\documentclass[aps,prb,floats,twocolumn,10pt,floatfix]{revtex4-1}

\usepackage{graphicx}
\usepackage{amssymb}
\usepackage{amsmath}

\usepackage{color}
\usepackage[dvipsnames]{xcolor}
\usepackage{slantsc}
\usepackage{float}
\graphicspath{{./FIG_FINALI/}}
\usepackage{bbold}
\usepackage[caption=false]{subfig}
\usepackage{array}
\usepackage{braket}
\usepackage{scrextend}
\usepackage{hyperref}
\hypersetup{backref,pdfpagemode=FullScreen,colorlinks=true,allcolors=blue}
\usepackage{leftidx}

\begin{document}

\preprint{APS/123-QED}
\title{Machine learning density functional theory for the Hubbard model}
\author{James Nelson}
\email{janelson@tcd.ie}
\author{Rajarshi Tiwari}
\email{tiwarir@tcd.ie}
\author{Stefano Sanvito}%
\email{stefano.sanvito@tcd.ie}
\affiliation{%
School of Physics, 
AMBER and CRANN Institute, 
Trinity College, 
Dublin 2, 
Ireland
}%
\date{\today}

\begin{abstract}
The solution of complex many-body lattice models can often be found by defining an energy functional 
of the relevant density of the problem. For instance, in the case of the Hubbard model the spin-resolved site 
occupation is enough to describe the system total energy. Similarly to standard density functional theory, 
however, the exact functional is unknown and suitable approximations need to be formulated. By using a
deep-learning neural network trained on exact-diagonalization results we demonstrate that one can construct 
an exact functional for the Hubbard model. In particular, we show that the neural network returns a ground-state 
energy numerically indistinguishable from that obtained by exact diagonalization and, most importantly, that 
the functional satisfies the two Hohenberg-Kohn theorems: for a given ground-state density it yields the external 
potential and it is fully variational in the site occupation.
\end{abstract}

\maketitle

Density functional theory (DFT)~\cite{DFT} is today the most widely used method for computing the electronic structure
of solids and molecules and it finds widespread applications in physics, chemistry, biology and materials science. 
The success of DFT has to be attributed to its solid theoretical foundation, contained in the Hohenberg-Kohn theorems
\cite{HK}, in a clear pathway to a practical implementation, formulated in the Kohn-Sham equations~\cite{KS}, 
and to the possibility of constructing a hierarchical ladder of approximations for the energy functional~\cite{Perdew}.
Although DFT has been developed to solve the problem of $N$ electrons and $M$ nuclei interacting through the 
long-range Coulomb potential, one can formulate DFT-type approaches also for many-body lattice 
models~\cite{HubbardDFT}. In this case the electron density is replaced by an appropriate density defined on the lattice, 
which becomes the fundamental quantity of the theory. For instance, the local on-site occupation, $\{n_{i\sigma}\}$, is the
relevant density of the Hubbard model. This is the expectation value of the number operator, 
$\hat{n}_{i\sigma}=c^\dagger_{i\sigma}c_{i\sigma}$, with $c^\dagger_{i\sigma}$ ($c_{i\sigma}$) being the fermionic 
creation (annihilation) operator at site $i$ for spin $\sigma=\uparrow,\:\downarrow$. In the lattice DFT framework the 
equivalents of the two Hohenberg-Kohn theorems can be demonstrated. 

The interest in lattice DFT is twofold. On the one hand, it provides a scalable numerical platform to investigate strongly correlated 
inhomogeneous systems as they approach the thermodynamic limit~\cite{Alcaraz2007,Saubanere2011}. On the other hand, 
it allows one to explore fundamental questions common to any density functional theory, such as the choice of the reference system 
in the construction of the energy functional~\cite{Lima2003}, the origin of the Mott gap~\cite{Lima2002}, the effects of strong 
correlation in quantum transport~\cite{Stefanucci2011,Vettchinkina2013,Pertsova2013}, the extension to the time
domain~\cite{Verdozzi2008,Karlsson2011,Akande2012}. For some lattice models exact solutions exist. These typical concern
the homogeneous case~\cite{lieb1968}, but they do not generalise to the inhomogeneous one~\cite{essler2005}. Importantly,
lattice DFT offers the ideal theoretical framework to treat on the same footing both the homogeneous and inhomogeneous
problem, as we are now going to explain.

Let us consider, as an example, the standard one-dimensional Hubbard model. The Hamiltonian operator writes
\begin{equation}
\hat{H}_U=\hat{T}+\hat{U}+\sum_{i\sigma} \hat{n}_{i\sigma}v_i\:,
\end{equation}
where $\hat{T}=-t\sum_{i\sigma}(\hat{c}_{i+1,\sigma}^\dagger \hat{c}_{i\sigma} + \hat{c}_{i\sigma}^\dagger \hat{c}_{i+1,\sigma})$
is the kinetic energy with hopping parameter $t$, $\hat{U}=U\sum_{i=1}\hat{n}_{i\uparrow}\hat{n}_{i\downarrow}$ is the Coulomb
repulsion of strength $U>0$, and the last term describes the interaction with an external potential $\{v_i\}$. If $v_i=v_0$ for every 
site $i$, one has the homogeneous case. The first Hohenberg-Kohn theorem establishes the existence of an energy functional,
\begin{eqnarray}\label{Efunctional}
E[\{n_{i\sigma}\}]=F[\{n_{i\sigma}\}]+\sum_{i\sigma}n_{i\sigma}v_i\:, \\
F[\{n_{i\sigma}\}]=\langle\Psi|\hat{T}+\hat{U}|\Psi\rangle\:,
\end{eqnarray}
where $F[\{n_{i\sigma}\}]$ is universal and independent from the external potential, $v_i$. This means that there is a one-to-one
correspondence between the density and the external potential, namely the knowledge of the former is enough to uniquely determine
the latter. Since the functional is universal, lattice DFT can be applied identically to both the homogeneous and the inhomogeneous 
problem. Note that for this lattice model $F[\{n_{i\sigma}\}]$ is universal only for a given $\hat{T}$ and $\hat{U}$ operator. This has
some important consequences. For instance, even in one dimension, arranging the sites in a ring or in a chain geometry yields 
two different $\hat{T}$'s. This means that $F[\{n_{i\sigma}\}]$ for a ring is different from $F[\{n_{i\sigma}\}]$ for a chain, even if both 
$t$ and $U$ remain the same. Finally the second Hohenberg-Kohn theorem guarantees that the energy functional is minimised at 
the ground-state density, $\{n_{i\sigma}^\mathrm{GS}\}$, to yield the ground-state energy, $E^\mathrm{GS}$, 
namely $E[\{n_{i\sigma}\}]\ge E[\{n_{i\sigma}^\mathrm{GS}\}]=E^\mathrm{GS}$.

As for standard DFT even in the lattice case the exact functional is not known and approximations need to be formulated. These
vary depending on the specific problem and usually proceed by interpolation from know exact solutions. Here we take a completely 
different approach and we construct an {\it exact} functional by using a machine learning model trained on exact diagonalization 
results. Machine learning (ML) models form a large class of algorithms, which have been traditionally used for data processing 
and analysis. Recently these numerical techniques have found a common place in many-body problems~\cite{Carleo2017}, 
phase transitions~\cite{Carrasquilla2017,Chng2017,Broecker2017}, Green's functions calculation~\cite{Arsenault2014}, and in 
the solution of single-particle problems in an arbitrary potential~\cite{Mills2017}. 

``Supervised'' ML algorithms construct a one-to-one correspondence between vectors or between vectors and 
scalars. We can then formulate the search for the energy functional as a ML problem, namely we can search for 
a non-analytic function, $f_\mathrm{ML}$, which associates to a given density (the vector of site occupation 
$\{n_{i\sigma}\}$) the corresponding energy. In particular our strategy consists in solving exactly the many-body 
problem for a set of different randomly chosen external potentials, $\{v_i\}$, and then to use the calculated
ground-state density and ground-state energy to define the ML model. In practice we compute
\begin{equation}\label{MLFunctional}
f_\mathrm{ML}:\:\{n_{i\sigma}^\mathrm{GS}\}\rightarrow E^\mathrm{GS}-\sum_{i\sigma}n_{i\sigma}^\mathrm{GS}v_i
=\langle\Psi^\mathrm{GS}|\hat{T}+\hat{U}|\Psi^\mathrm{GS}\rangle\:,
\end{equation}
where $|\Psi^\mathrm{GS}\rangle$ is the many-body ground-state wave function. The fact that we use a set of 
random external potentials effectively allows us to explore a broad range of densities, and hence to map accurately
the functional.

We apply our strategy to the one-dimensional Hubbard model for a system of $L=8$ sites arranged in a ring geometry
(periodic boundary conditions). Furthermore we restrict ourself to the paramagnetic quarter filling case, where the
the total number of electrons is $N=4$ and we have $N_\uparrow=N_\downarrow=2$. In order to fully determine the
model we set~\cite{foot} $U=4$ and $t=1$ and, as mention before, we construct the dataset for the ML model by exact 
diagonalization~\cite{sharma2015organization}. The random external potential is taken according to a uniform distribution 
with $v_i\in [-W,+W]$. In particular we have constructed several external potential distributions with $W$ varying between
0.005$t$ and 2.5$t$. Furthermore, in order to prevent the dataset from having large
fluctuations in the total energy we neglect potentials yielding to total energies 0.15$t$ larger than that of the homogeneous 
case. Note that the $SU(2)$ symmetry of the problem and the condition $N_\uparrow=N_\downarrow$ also guarantee 
that the local site occupation remains spin unpolarised, namely we have $n_{i\uparrow}=n_{i\downarrow}$. 
This means that in the special case investigated here the functional depends only on one of the two spin densities, 
for instance on $\{n_{i\uparrow}\}$.

\begin{figure}[h]
\includegraphics[width=1.0\linewidth]{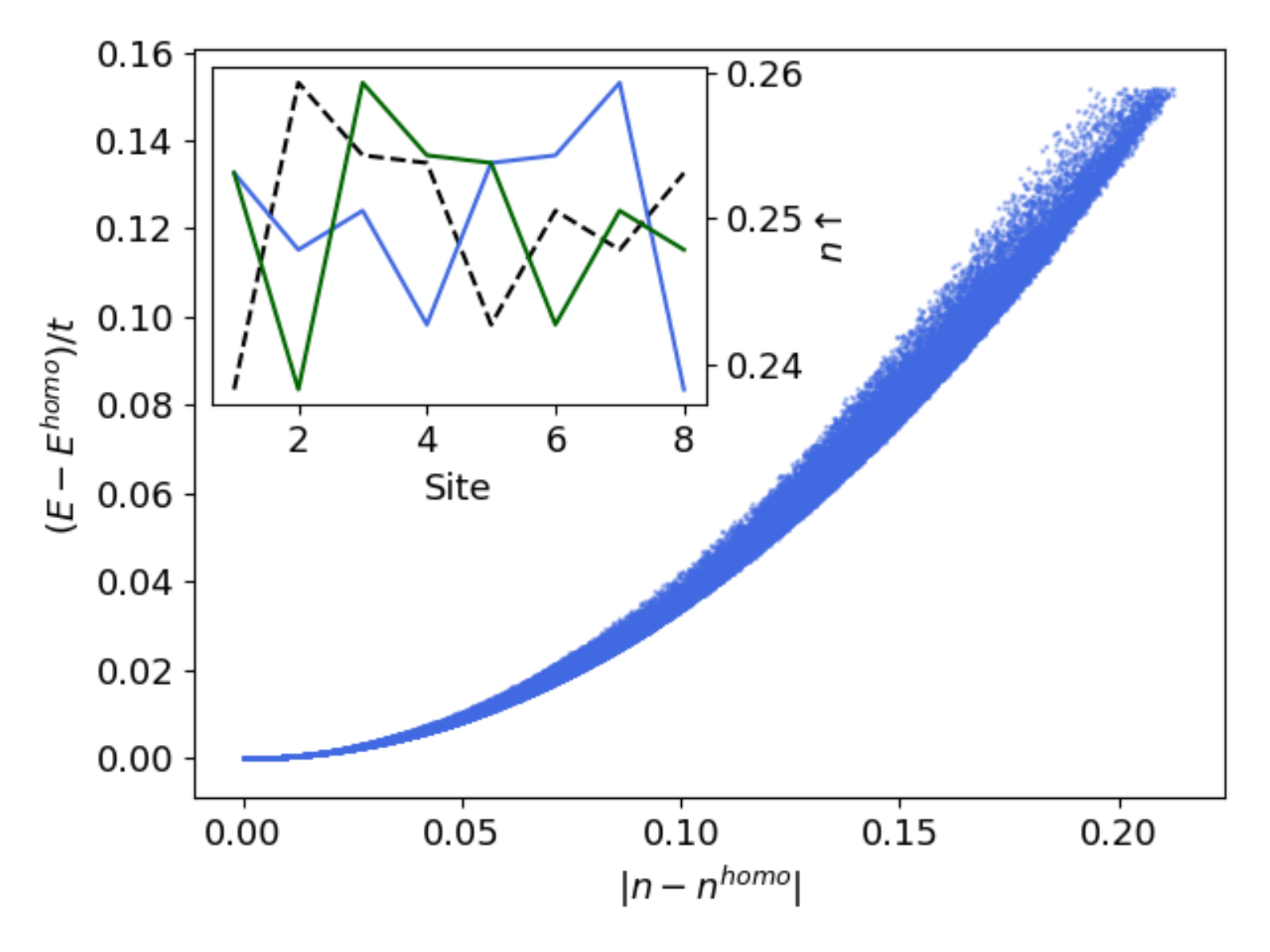}
\captionsetup{justification=justified,singlelinecheck=false}
\caption{(Color online) Structure of the data used to construct the ML model. We plot the ground-state total energy 
as a function of the euclidean distance of the corresponding charge density with respect to that of the homogeneous 
case, $n_{i\sigma}=1/4$, for a large selection of external potentials, $\{v_i\}$. The total energy is measured with respect 
of that of the $v_i=0$ case, which corresponds to $E^\mathrm{homo}=F[\{1/4\}]$. Note that the data points in the main plot
are those generated numerically and do not consider the symmetries of the system. In the insert we show examples of 
charge densities constructed from a given one (dashed line) by applying the symmetry operations of the system: 
mirror symmetry (blue line) and the translational symmetry (green line).}
\label{fig:landscape}
\end{figure}

In order to increase the size of the dataset without performing further exact diagonalization steps we include
configurations obtained from the calculated ones by applying the allowed symmetry operations. In particular for any potential
$v_i$ the mirror-symmetric potential ${v_{i}\rightarrow v_{L+1-i}}$ yields a mirror-symmetric charge density with 
identical total energy. A similar situation applies to potentials obtained by translation, namely ${v_{i}\rightarrow v_{i+1}}$.
Examples of the charge density profiles obtained with such symmetry operations are presented in the inset of
Fig.~\ref{fig:landscape}. The addition of such configurations drastically improves the ML model. The improvements 
has two main origins. On the one hand the dataset is larger. On the other hand the inclusion of the degenerate
configurations allows the model to learn about the symmetries of the system. 

In the construction of the ML model the dataset is split into four mutually exclusive subsets. The training set, 
containing 52,500 samples, is used to train the model. The validation set, containing 26250 samples, is employed 
to select the best ML model. The test set, also containing 26250 samples, serves to estimate the generalization 
error of the model, namely how well it performs on new data never seen before. Finally we set aside 100 configurations 
to test the gradient descent scheme for the demonstration of the second Hohenberg-Kohn theorem. These latter 
configurations are chosen uniformly over energy such that the entire range is explored. The structure of our dataset 
is presented in Fig.~\ref{fig:landscape}. In the figure we plot the ground-state total energy as a function of the 
euclidean distance of the corresponding charge density with respect to that of the homogeneous case, $n_{i\sigma}=1/4$. 
Larger euclidean distances are associated to the choices of external potential with larger fluctuations. As expected,
the distribution of total energies gets broader as the deviation from the homogenous case gets larger. Note that
the homogeneous case, $v_i=0$, is associated to the minimum of the distribution, where we have 
$E^\mathrm{GS}=F[\{1/4\}]$.

After having tested a few ML models we have opted for a convolutional neural network, which is found to perform
better than standard neural networks~\cite{goodfellow2016deep}. All the machine learning algorithms have 
been implemented by using the Keras python package \cite{chollet2015keras}. In the convolutional neural network
in order to fully capture all the information we extend each of the occupation vectors, by adding their first $k$ components 
on to the end of the vector, thus creating a $L+k$ component-long vector. In doing so $k$ is the size of the one-dimenisonal 
convolution window, in our particular case $k=3$. This procedure allows us to encode the interaction between the sites at
the end and the sites at the beginning of the vector. The convolution neural network used has 8 convolutional filters, 
followed by 2 fully connected layers each with 128 units and finally an output layer.

\begin{figure}[h]
\centering\includegraphics[width=0.9\linewidth]{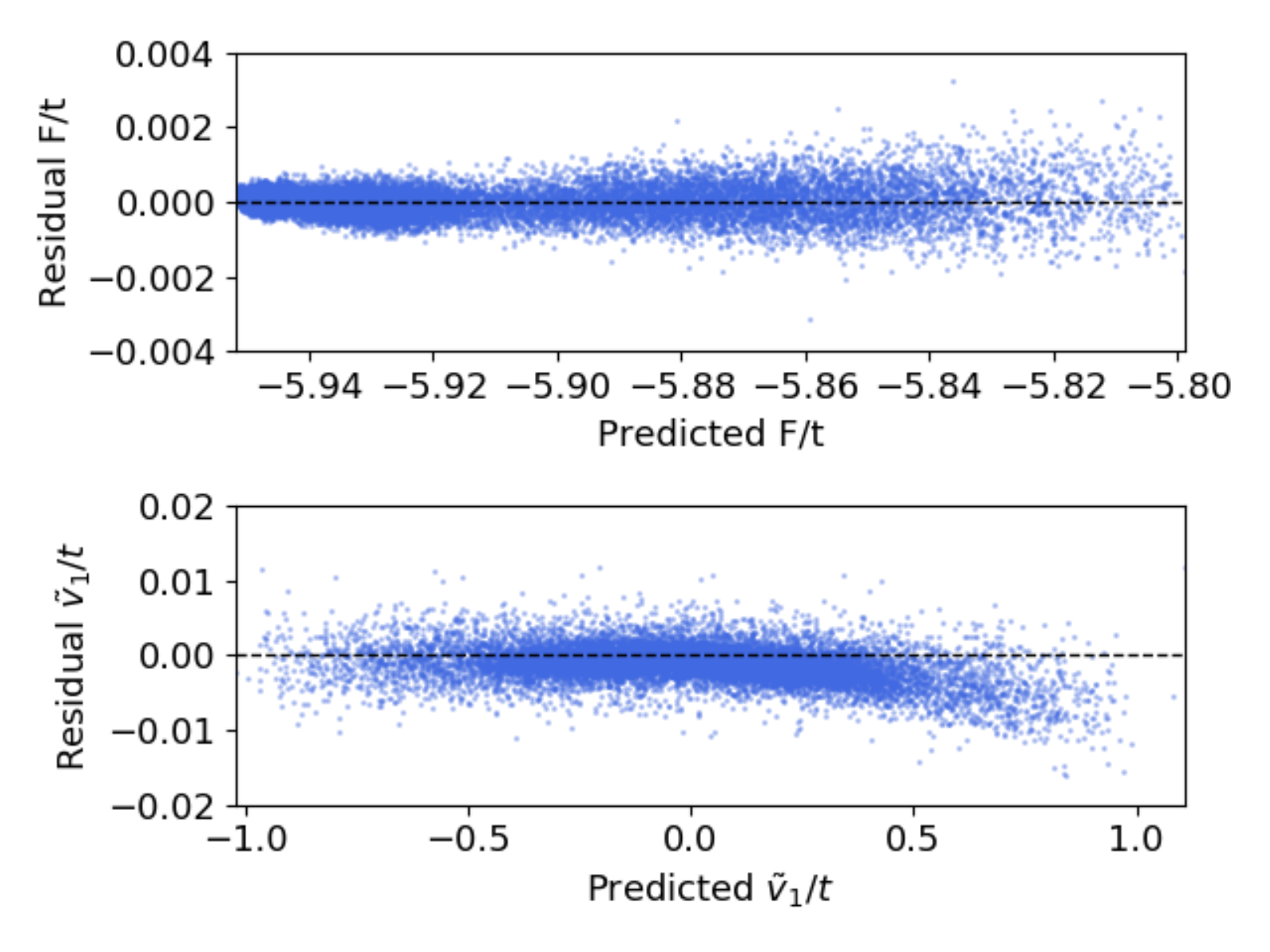}
\caption{Numerical accuracy of the ML model measured by the residuals, namely the difference between a predicted quantity 
and its exact value. In the upper panel we show the residual of $f_\mathrm{ML}$ as a function of the corresponding 
exact-diagonalization result for the entire test set. The model is almost indistinguishable from the exact results, 
demonstrating that $f_\mathrm{ML}$ is the exact functional. The mean absolutely error is $0.0002t$, namely it is about 
0.02\%.
In the lower panel we present a numerical demonstration of the first Hohenberg-Kohn theorem. The residual of the external 
potential at site $i=2$, measured with respect to that at site $i=1$, $\tilde{v}_1=v_2-v_1$, is plotted against the exact value 
for the entire test set. The mean absolute error is $0.001t$. Similar curves can be obtained for the other on-site energies.}
\label{fig:universal}
\end{figure}

The accuracy of our ML model can be appreciated by looking at the upper panel of Fig.~ \ref{fig:universal}, where we present 
the residual of the predicted $F[\{n_{i\sigma}^\mathrm{GS}\}]$, namely $F^\mathrm{GS}_\mathrm{exact}-f_\mathrm{ML}$, 
calculated over the test set against the predicted results. From the figure it can be clearly appreciated that the ML 
model is almost indistinguishable from the exact functional. Its mean error over the entire test set is in fact 0.0002$t$, namely it is 
of the order of 0.02\%. This remarkable accuracy confirms that the functional has been fully learned by our ML model. 

We now proceed to demonstrate that the ML functional, $f_\mathrm{ML}$, satisfies both the Hohenberg-Kohn theorems. 
The first Hohenberg-Kohn theorem states that the ground state density uniquely determines the external potential up 
to a constant \cite{HK}. In order to numerically demonstrate the theorem we take the convolutional neural network
constructed before and we generate as an output a seven-component vector. This contains the external potential, 
namely the on-site energies $\{v_i\}$, which are measured by imposing that the potential of the first site is $v_1=0$. 
Such constrain ensures that that all the potentials are determined with the same constant. In other words our 
constant-rescaled external potential is the vector
\begin{equation}
\tilde{\textbf{v}} = (v_2-v_1, v_3-v_1,...,v_L-v_1)\:.
\end{equation}

The numerical proof of the first Hohenberg-Kohn theorem is presented in the lower panel of Fig.~\ref{fig:universal}, where we compare 
the residual of the first component of $\tilde{\textbf{v}}$, namely $\tilde{v}_1=v_2-v_1$, against the exact results. 
As before the agreement over the entire test set is nearly perfect with a mean absolute error of $0.001t$. Similar 
agreements are found for the other on-site energies, so that one has to conclude that 0.001$t$ is the precision of our ML 
model to determine the external potential.

Finally we proceed to demonstrate the second Hohenberg-Kohn theorem, namely we show that the ground-state energy
can be found as the minimum of the universal functional at the ground-state density. Strictly speaking this information was 
not explicitly used in the construction of our ML functional, since $f_\mathrm{ML}$ interpolates only at the ground state 
and not around it. However, since our data set includes a vast number of external potentials, it includes a vast number of 
different densities. As such $f_\mathrm{ML}$ has been {\it de fact} constructed by extensively exploring the entire 
density landscape. The minimization of the energy functional (\ref{Efunctional}) is carried out by gradient 
descent~\cite{goodfellow2016deep}. The generic derivative of $E[\{n_{i\sigma}\}]$ writes 
\begin{equation}
\frac{\partial E}{\partial n_{j\sigma^\prime}} = v_j + \frac{\partial}{\partial n_{j\sigma^\prime}} F[\{n_{i\sigma}\}]\:,
\end{equation}
where the gradient of the ML model, $F[\{n_{i\sigma}\}]=f_\mathrm{ML}$, is estimated by using second order finite-differences. 
The search for the minimum must also satisfy two conditions, namely ${0 \le n_{i\sigma} \le 1}$ and particle conservation, 
${\sum_{i=1}^L n_{i\sigma} = N_\sigma}$. The first condition is imposed by simply halting the gradient descent algorithm, 
whenever it is violated. In contrast the second one is enforced by normalizing the occupations after each update.
\begin{figure}[h]
\centering\includegraphics[width=0.9\linewidth]{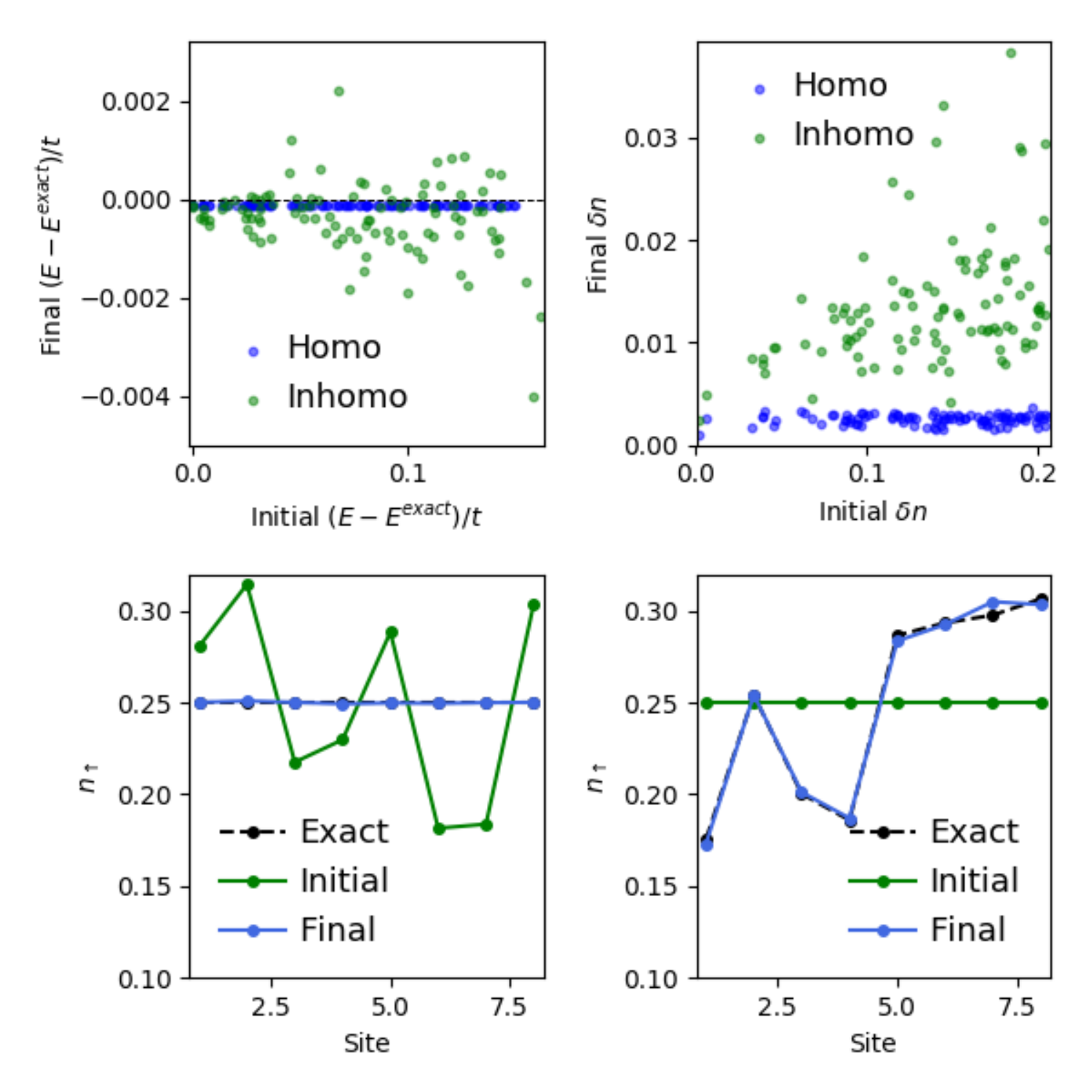}
\caption{Numerical demonstration of the second Hohenberg-Kohn theorem. The top two panels show the error in the final 
converged energy (left)  and density (right) [see Eq.~(\ref{nmetric})] as a function of the distance of the initial quantity from 
the exact one. Note that when searching for the homogeneous density/energy (blue dots) the gradient descent algorithm
appears more accurate than when looking for the inhomogeneous one (green dots).
The two lower panels show examples of the converged occupations, when searching for the ground state density of either 
the homogeneous (left) or the inhomogeneous (right) system.}
\label{2ndHK}
\end{figure}

The accuracy of the computed site occupations is measured as 
\begin{equation}\label{nmetric}
\delta n = \sqrt{\sum_{i\sigma}|n_{i\sigma}-{n}_{i\sigma}^\mathrm{exact}|^2}=
\sqrt{2}\sqrt{\sum_{i}|n_{i\uparrow}-{n}_{i\uparrow}^\mathrm{exact}|^2}\:,
\end{equation}
where ${n}_{i\sigma}^\mathrm{exact}$ is the exact site occupation and the second equality follows from 
$n_{i\uparrow}=n_{i\downarrow}$. Similarly we compute the difference in energy from the reference system.
In particular we perform two tests. In the first one we set the external potential to $v_i=0$ and search for 
$\{n_{i\sigma}\}$ and $F[\{n_{i\sigma}\}]$ corresponding to the homogeneous case starting from a random 
non-homogeneus occupation. In the second one we start from the homogeneous occupation and test the 
convergence to $\{n_{i\sigma}^\mathrm{GS}\}$ corresponding to the inhomogeneous potential $v_i\ne0$. 
Such exercise is performed over the gradient descent set (100 samples) and the results are presented in 
Fig.~\ref{2ndHK}. 

The two top panels demonstrate that $f_\mathrm{ML}$ is variational. In these we show the difference in the
energy (left-hand side panel) and site occupation (right-hand side panel) between the expected exact values
and that found by gradient descent, with respect to their initial value. This effectively explores how well 
$f_\mathrm{ML}$ has learned about the energy functional landscape, and looks at how close it converges to the 
ground-state energy and occupation as a function of how far the initial occupation/energy was. We use blue dots 
when searching for the ground state of the homogeneous case and green ones for that of the inhomogeneous. 
If $f_\mathrm{ML}$ is exact all dots will be on an horizontal straight line at zero.  In general we find
that the ground-state quantities (energy and density) are reached within a few percent regardless of their initial value. 
When searching for the ground state of the homogenous system the error is minimal and almost independent from 
the initial condition. In contrast the search for $\{n_{i\sigma}^\mathrm{GS}\}$ and $E^\mathrm{GS}$ corresponding 
to an inhomogeneous potential is less accurate and the average error grows linearly with the distance from the 
initial occupation from the final one. 

The two lower panels instead display two examples of converged site occupation. The left-hand side panel show
that one can recover the homogeneous occupation when the gradient descent starts from an inhomogeneous 
one, while that on the right-hand side depicts the opposite, namely that an almost exact inhomogeneous 
occupation can be found by starting from the homogeneous one. 

In conclusion we have used machine learning to construct the exact energy functional for the inhomogeneous 
Hubbard model. This is provided by a convolutional neural network constructed over exact results obtained
by exact diagonalization. The functional appears numerically indistinguishable from the exact solutions and
satisfies both the Hohenberg-Kohn theorems, namely it establishes a one-to-one correspondence between
the site occupation and the external potential, and it is variational. Here the functional was constructed for the
quarter filling diamagnetic case, but the same procedure can be applied to any filling factors and number of sites, 
including the possibility to describe magnetic ground states. Our results demonstrate that ML can be
used to define exact density functional theories and offer the potential to be expanded to other lattice and
continuous models. 

\subsubsection*{Acknowledgments}

Funding is provided by the Irish Research Council (JN) and by the European Research Council project {\sc quest} (RT).
We acknowledge the DJEI/DES/SFI/HEA Irish Centre for High-End Computing (ICHEC) and Trinity Centre for High 
Performance Computing (TCHPC) for  the provision of computational resources.


\begin{thebibliography}{99}
\bibitem{foot}Throughput the paper we measure all energies in unit of $t$.
\end{thebibliography}
\end{document}